\documentclass[letterpaper]{article} % DO NOT CHANGE THIS
\usepackage{aaai2026}  % DO NOT CHANGE THIS
\usepackage{times}  % DO NOT CHANGE THIS
\usepackage{helvet}  % DO NOT CHANGE THIS
\usepackage{courier}  % DO NOT CHANGE THIS
\usepackage[hyphens]{url}  % DO NOT CHANGE THIS
\usepackage{graphicx} % DO NOT CHANGE THIS
\usepackage{natbib}  % DO NOT CHANGE THIS AND DO NOT ADD ANY OPTIONS TO IT
\usepackage{caption} % DO NOT CHANGE THIS AND DO NOT ADD ANY OPTIONS TO IT
\usepackage{algorithm}
\usepackage{algorithmic}

\usepackage{newfloat}
\usepackage{listings}
\DeclareCaptionStyle{ruled}{labelfont=normalfont,labelsep=colon,strut=off} % DO NOT CHANGE THIS
\floatstyle{ruled}
\newfloat{listing}{tb}{lst}{}
\floatname{listing}{Listing}
\usepackage{tabularx}

\title{Culturally Situated AI Safety for Youth: Saudi Arabian Perspectives of
Youth, Parents and Teachers}
\author{
    Aljawharah Alzahrani\textsuperscript{\rm 1},
    Tanusree Sharma\textsuperscript{\rm 1}
}
\affiliations{
    \textsuperscript{\rm 1}College of Information Sciences and Technology, Pennsylvania State University\\
    ama9555@psu.edu, tanusree.sharma@psu.edu
}

\usepackage{bibentry}
\begin{document}

\maketitle

\begin{abstract}
Generative AI (GenAI) tools are widely used by youth and have introduced new privacy and safety challenges. While prior research has explored youth's safety in GenAI within a Western context, it often overlooks the cultural, religious, and social dimensions of technology use that strongly shape youth’s digital experiences in countries like Saudi Arabia. 
To address this gap, this study explores youth's (aged 7-17), parents' and teachers' interactions with GenAI tools and risk perceptions through a non-Western lens. We  analyzed 736 Reddit psots, and 1,262 X (Twitter) posts, and conducted interviews with 31 Saudi Arabian participants (8 youth, 13 parents, 10 teachers). 

Our findings highlight context-dependent and relational privacy and safety needs of GenAI use from non-Western context, which are often shaped by communal structure and prescribed norms. We found significant risks tied to youth’s disclosure of personal and family information, which conflict with culturally rooted expectations of modesty, privacy, and honor, particularly when youth seek emotional support from GenAI. These risks are further compounded by socio-economic factors such as cost-saving practices leading to the use of shared GenAI accounts (e.g., ChatGPT) within families or even among strangers. 
We provide design implications reflecting parents’ and teachers’ expectations of how youth should use GenAI. This work lays the groundwork for inclusive, context-sensitive parental controls that adhere to cultural norms and values.
\end{abstract}


\section{Introduction}

Youth are increasingly interacting with Generative AI (GenAI) from an early age, which shapes how they learn, play, and communicate. Generative Artificial Intelligence (GenAI) tools such as ChatGPT, Gemini, and DeepSeek have become widely accessible and are rapidly transforming these experiences.
%Generative Artificial Intelligence (GenAI) tools such as ChatGPT, Gemini, and DeepSeek have become widely accessible and are rapidly transforming how people learn, socialize, and play. 
Tools such as ChatGPT, which supports advanced text generation for content creation, and DALL·E, which produces highly realistic images from textual prompts, highlight the transformative role GenAI plays in digital media and design~\cite{10.1145/3613904.3642794}. Meanwhile, platforms like Character.ai expand the scope of interaction by offering character-based AI chatbots, enabling personalized and dynamic conversations that often mirror human dialogue ~\cite{10.1145/3698061.3726934, zhang2025riseaicompanionshumanchatbot}. While the global uptake of GenAI tools has been remarkable, adoption in Saudi Arabia has been especially significant, with a recent survey reporting that 80\% of adults in Saudi Arabia actively engage with GenAI, which surpasses usage rates in the United States \cite{Alhamawi_2025}. 

This momentum is particularly visible among younger audiences, raising critical questions about how these technologies intersect with relational and collective privacy embedded within cultural values. In Saudi Arabia, youthand young adults aged 10–19 constitute the second-largest demographic of AI users at 26.4\%, closely following adults aged 20–29 at 27.3\%, with ChatGPT currently ranking as the most downloaded AI application in the country~\cite{CST_2024}. Such widespread engagement signals that youth are not only aware of GenAI but actively engage with it, exposing them to unique risks associated with their developmental stage and psychological vulnerabilities~\cite{neugnot2024future}. These risks can be further compounded by the cultural context, where family structures and communal expectations influence both usage patterns and the potential for harm, including breaches of privacy, exposure to inappropriate content, and conflict with culturally prescribed norms~\cite{abokhodair2017privacy, farooq2024exploring}.
%their learning, socialization, and cultural development since 

Existing research has explored youth’s interactions with GenAI in Western contexts, highlighting risks such as inappropriate content, over-reliance on AI, and misinformation~\cite{yu2024exploring, yu2025understanding}. Yet, these studies often overlook the cultural, religious, and social dimensions of technology use that strongly shape youth’s digital experiences in non-Western contexts. In Saudi Arabia, discussions around technology are not only framed in terms of individual safety but also in terms of communal trust, family honor, and adherence to religious norms~\cite{alashwali2022saudi, alsiyat2025smart, mian2023investigating}.
%This presents unique risks and expectations for children’s use of GenAI. 
Evidently, previous research across digital domains in Saudi Arabia, such as youths’ use of social media platforms like TikTok and Snapchat, also highlights concerns regarding privacy, surveillance, and the erosion of family values~\cite{alashwali2022saudi, aleisa2020privacy, alqarni2024iot}. In particular, data collection by foreign companies has raised broader questions related to national sovereignty and identity~\cite{alorini2025unplug, saudigazette2019online}. %Studies on IoT and smart home technologies document mistrust toward devices that  might monitor household activities, 
Privacy in Saudi Arabia is not just about individual data security; it is intertwined with broader questions of religion, morality, and family cohesion. Given GenAI’s conversational nature, and potential to mediate sensitive topics, such as religion or personal relationships, as demonstrated in Western contexts, these concerns may be amplified in non-Western settings, where cultural differences may intensify sensitivities.
%reflecting cultural emphasis on the sanctity of private life. 

To address this gap, we examined two research questions from three user groups with different perceptions of GenAI from a non-Western context:
%\\\textbf{RQ1} How do youth, parents, and teachers from Saudi Arabia engage with GenAI tools?

\textbf{RQ1} What are the current strategies for ensuring safety of youth using GenAI in Saudi Arabia? 

%\tanusree{for this, we will need a page of regualtion analysis of saudi arabia}
\textbf{RQ2} What unique risks emerge from youths’ interactions with GenAI in the Saudi Arabian context? 
%and how thye compound harm through different interaction pathways?
%What concerns do parents and teachers from Saudi Arabia have about children's usage of GenAI tools?

%\textbf{RQ3} What are the mediation strategies employed by parents and teachers to ensure youth’s safety with GenAI?

%What is the expected safe use of GenAI suggested by Saudi parents and teachers?

To answer these two research questions,
%we employed a mixed-methods approach. We analyzed 736 Reddit posts and 1,262 X (formerly Twitter) posts to examine Saudi users’ knowledge, attitudes, and perceptions regarding GenAI tools, followed by a
we conducted interviews with 8 youth, 13 parents, and 10 teachers. We also analyzed 736 Reddit posts and 1,262 X (formerly Twitter) posts to identify the risks associated with youth's interactions with GenAI, and to explore the strategies employed by parents and teachers to mediate and mitigate potential harms.

\textbf{Main Findings.} 
%\tanusree{add main findings in 5 lines} 
%We found that youth’s use of GenAI extended well beyond educational support. They engaged with these tools to explore religious questions, seek romantic companionship, and even disclose sensitive family matters. In contrast, parents and teachers used GenAI to augment tutoring and personalized educational support, sometimes unintentionally revealing information about their children.
%From a cultural lens, children’s inquiries into religious topics emerged as a unique risk, as Parents and teachers feared that inaccurate or misleading AI-generated responses could distort children’s beliefs. 
%Parents also expressed concern about GenAI’s influence on family relationships, particularly when children reported turning to ChatGPT as a confidant more often than to their families.
By synthesizing our findings:
\begin{itemize}
%\begin{enumerate}[leftmargin=*,labelsep=0.4em,
                % itemsep=1pt,parsep=0pt,
                 % topsep=2pt,partopsep=0pt]
\item 
%We expanded the existing Western-centric youth-GenAI risk taxonomy \cite{yu2025understanding}
Our culturally situated taxonomy introduces three unique risk categories: \textit{Religious and Moral Risk}, \textit{Cultural Norms Violation Risk}, and \textit{Data Privacy Violation Risk}, which have not been addressed in prior AI risk taxonomies \cite{yu2025understanding,critch2023tasra,slattery2024ai}.

\item We identified significant risks associated with youth disclosing personal and family information, details of romantic interest or family conflict, which often violate culturally rooted expectations of modesty, privacy, and family honor.
%in ways different from general privacy concerns, especially when youth turn to GenAI for emotional support. S
Similarly, youth's exposure to GenAI's inconsistent religious interpretations
%with local norms 
reflects concerns related to religious appropriateness rather than universal content safety.

\item Compounded risk by socio-economic factors, such as cost-saving practices leading to shared GenAI accounts (e.g., ChatGPT) within families or with strangers.
\end{itemize}

\section{Related Work}

\subsection{Security and Safety for Youth in Generative AI}

The rapid adoption of GenAI tools by youth has opened up many great opportunities but also significant concerns.  GenAI and its impact on youth's lives have been explored widely in Western contexts. Prior work has documented youth's interactions with GenAI tools from various aspects, such as integrating GenAI in youth's education, enhancing youth creativity with GenAI, youth safety in GenAI, and youth GenAI risks taxonomies \cite{yu2024exploring,yu2025understanding}, and mediation of GenAI \cite{zhang2023sa,yu2024exploring}. Prior research showed key risks in youth's GenAI interactions, such as relying on GenAI for emotional and mental support \cite{yu2024exploring,yu2025understanding}, over-reliance on GenAI \cite{yu2024exploring,yu2025understanding}, GenAI bias and discrimination \cite{yu2025understanding}, and privacy concerns \cite{yu2024exploring}. Parents and families play a critical role in mediating and managing GenAI's risks for youth \cite{yu2024exploring,zhang2023sa}; however, evidence shows that parents often struggle to assess the safety of GenAI tools for youth\cite{yu2025understanding}.

Several studies have documented risks related to youth's online privacy and security \cite{williams2023youth,sun2021child,cao2024understanding,livingstone2014their}. Prior work \cite{williams2023youth} examined how youth understand and manage online privacy and security, and found that children as young as six understand the importance of privacy; however, their understanding is limited to interpersonal risks (e.g., ``stranger danger") rather than institutional or commercial data practices. The study also revealed that parents rely heavily on restrictive monitoring and limiting use rather than having conversations about nuanced security and privacy knowledge \cite{williams2023youth}. Beyond privacy, misinformation and generated content have emerged as security concerns for youth. Prior work documented youth's and parents' experiences with misinformation on social media platforms (e.g., TikTok and YouTube) \cite{sharevski2024children}. These studies highlight that existing work has largely developed in Western contexts and may not translate effectively to non-Western settings, calling for culturally appropriate approaches to GenAI safety.     

While recent work has begun to examine youth's interactions with GenAI tools in non-Western contexts, such as Brazil \cite{fortim2025brazil}, Kenya \cite{kaberi2025kenya}, and India \cite{kasturi2025india}, youth's use of GenAI in Saudi Arabian context specifically remains under-researched. Moreover, insights from prior work on youth’s privacy and safety concerns in Western contexts may not fully translate to parental controls design for Middle Eastern users, given differences in cultural norms, regulatory environments, parental mediation styles \cite{Dwairy2006}, and socio-technical infrastructure. For example, in Western contexts, privacy concerns may focus on individual rights, while in Middle Eastern or Asian contexts, family and community norms and government surveillance may play larger roles.
Moreover, cultural factors shape the way people interact, their values, and consequentially, how they use the internet \cite{Moubarak2024RiskAO}. Therefore, culture shapes how youth deal with online risks. For example, studies show that cultural barriers hinder youth from admitting to being exposed to online threats or inappropriate content and from reporting them to their parents, for fear to defying these barriers \cite{Alqahtani2017InternetRF,Teimouri2016AssessingTV}. Saudi Arabia is a conservative society that has holds these cultural norms, where parents have expectations and trust that their youth would not see or engage in discussions about sensitive or taboo topics such as inappropriate content on the Internet. 
% \vspace{-2mm}
\subsection{Digital Safety of Youth in Middle Eastern Countries}
% \vspace{-2mm}
Although, in 2022, a Boston Consulting Group (BCG) study \cite{BCG} reported that youth in the Middle East and Latin American countries experienced the highest levels of online threats compared to youth in North America and Europe, research on youth’s online safety in the Middle East region is limited. The BCG's study \cite{BCG} reported that the online threats youth are exposed to are numerous and come in many forms, such as pop-up ads, inappropriate content, cyberbullying and harassment, sexual approaches, malware and phishing. Few Middle Eastern studies have documented these online risks that youth faced, such as cyberbullying \cite{MohamedOthman2025UnravellingCA,Rehima2020TheFO,Farooq2023PrevalenceOC,Alfakeh2021ParentsPO} and online extortion \cite{Martin2023FromET}.

Youth's digital literacy refers to the knowledge and skills that allow youth to navigate the digital world, both effectively and safely \cite{UNICEF}. Youth's cybersecurity education helps youth protect themselves by recognizing online threats \cite{Ebrahimi2025CybersecurityEE}. Youth's AI literacy is defined as the set of skills that enables youth to understand how AI works and how to use it \cite{Su2023ArtificialI}. These are important skills for Middle Eastern youth to acquire in order to protect themselves from online risks. However, in Arab nations, few studies have examined youth's digital literacy \cite{Almethen2024TheRO} and cybersecurity education\cite{AlShabibi2021CybersecurityAA}. To the best of our knowledge, there is no study that has examined youth's AI literacy nor GenAI risks in the Middle East region.
Moreover, a prior study reported that Saudi parents lacked parental mediation; they did not discuss online threats with their youth nor monitored their online activities. However, parents have expressed interest in using parental control apps to monitor their children's online use \cite{Alqahtani2017InternetRF}. This shows that there is a desire to protect their children, but they lack the knowledge to do so, as well as a lack of Arabic language resources to raise awareness about youth's online risks.

\subsection{Current AI Safety Tools in Western and Middle Eastern Countries}

%OpenAI the creator of ChatGPT and major tech companies (e.g. Google, Meta) pledged to commit to the child safety principle in developing and maintaining GAI models to prevent the rising threat of GAI in child exploitation\cite{The-independent}. In response to the recent lawsuit of a 16 years old parents claiming ChatGPT contributed to their son self-harm, OpenAI introduced new parental controls for ChatGPT where parents will be able to have more control over chats with ChatGPT  \cite{CNN,NBC}.

\textit{The current state of AI moderation tools:} OpenAI's new parental controls for ChatGPT have rolled out, which allows parents to link their own ChatGPT accounts with their teen's account (minimum age 13) using an email invitation, manage how ChatGPT interacts with their teenss, manage and disable features (e.g., memory and chat history), and receive notifications when the system detects that a "teen is in a moment of acute distress" \cite{OpenAI}.  Google's Gemini app availability for youth under 13 is controlled by Google's Family Link parental controls, which allow parents to control their children's access to the app and manage their child's Gemini settings \cite{theverge,Google}. Google stated that youth's data will not be collected to train AI \cite{theverge,Google2}. However, these AI safety tools are  not culturally adaptive and are limited in understanding Arabic content.

%\textit{Built-in parental tools from Apple and Google:} Apple's built-in primary parental controls on child's iPhone or iPad managed through the Screen Time feature, allowing parents to turn on Content and Privacy Restrictions to restrict inappropriate content, manage apps downloads and purchases and change privacy settings\cite{Apple}. Google's parental control is the Family Link app, allows parents to set screen time limits for app usage, block apps, manage content on Google services (e.g., YouTube, Chrome)  content, and locate child on the map \cite{Google2}. However, these safety tools work on deceive level and indifferent for GAI tools, not culturally adaptive, and youth can easily bypass restrictions.

\textit{The current state of regional and culturally-sensitive tools:} 
In 2025, the UAE's telecoms company (e\&) launched a new safety control feature that allows Emirati parents to filter content, block unwanted apps, block access to unsafe websites, and limit Internet use \cite{eand2025}. This parental control is designed specifically to meet the Emirati families needs. Moreover, the UAE Ministry of Possibilities and Snap Inc. developed a new parental control feature in Snapchat's Family Center; this feature allows parents to view youth's friends lists without viewing the actual conversations, in order to protect youth's privacy and to report abusive accounts \cite{ArabNews2}. In addition to these safety tools, there is a Saudi initiative called ``Qayyem" on Game consols (e.g., Xbox, Play Station) that aims to protect youth; it provides a rating system for popular games based on culturally-sensitive topics, such as ``conflicts with Islamic values" or ``decency, morality, and good behavior" \cite{alashwali2022saudi}.
However, these safety tools are limited to Internet use, device, or social media use, and there is an absence of region-specific AI safety tools tailored to youth’s needs. Therefore, we aim to address youth’s, parents’, and teachers’ needs for AI safety in these contexts.

\begin{table*}[t]
\centering
\small
\setlength{\tabcolsep}{2.5pt}

\begin{tabular}{@{}
p{2.75cm}
p{4.5cm}
p{4.5cm}
ccccc
@{}}
\hline
\textbf{Instrument} &
\textbf{Coverage} &
\textbf{Youth + GenAI Gaps} &
\textbf{D1} & \textbf{D2} & \textbf{D3} & \textbf{D4} & \textbf{D5} \\
\hline

Personal Data Protection Law (PDPL) &
Personal-data processing, including automated processing &
No dedicated youth's-data or synthetic-data provisions &
O & O & -- & -- & -- \\

Anti-Cyber Crime Law (2007) &
Unauthorized access, interception, privacy invasion, fraud, defamation, extortion, and harmful content &
No AI-generated/deepfake provisions or child-specific protections &
-- & O & O & -- & -- \\

SDAIA AI Ethics Principles (2023) &
Fairness, privacy, security, transparency, safety, accountability, and social benefit &
Non-binding; no youth- or age-specific safeguards &
O & O & O & -- & -- \\

SDAIA Generative AI Guidelines (2024) &
Responsible GenAI use, disclosure, deepfakes, oversight, risk management, and watermarking &
Non-binding; no child-safety-by-design provisions &
O & O & Y & -- & O \\

SDAIA AI Adoption Framework (2024) &
National framework for responsible AI adoption &
No consumer-GenAI or youth-specific safeguards &
O & O & O & -- & -- \\

Child Protection Law (2014) &
Protection of minors from abuse, neglect, sexual exploitation, and inappropriate content &
Predates GenAI; no digital, AI, platform, or developer duties &
-- & -- & -- & -- & -- \\

\hline
\end{tabular}

\caption{Saudi Arabian regulatory landscape mapped to GenAI child-safety dimensions.
Key: Y = direct; O = partial/indirect; -- = not addressed.
D1 = youth's data processing/consent; D2 = data privacy; D3 = AI-content/deepfake labeling;
D4 = synthetic CSAM; D5 = harmful-content exposure.}
\label{tab:saudi-regulation}
\end{table*}

\section{Regulatory Environment of Saudi Arabia}
%\tanusree{a page and a table of different regualtion in saudi-arabia in digital safety, ai safety, citizensafety, secuirty, etc,}

Saudi Arabia currently has no dedicated law regulating AI or GenAI, and no legislation addressing the safety of youth in GenAI environment \cite{nemko_ksa, hsf_ksa}. Instead, AI is governed through a layered framework in which binding data protection law works alongside non-binding policies \cite{nemko_ksa}. This analysis maps the existing and proposed laws and regulations that govern AI, digital safety, and cybersecurity in order to assess how far they protect youth from the specific risks of GenAI. 

The first layer consists of foundational binding data and cyber laws that govern the use of technology in general, such as the Personal Data Protection Law (PDPL) and the Anti-Cyber Crime Law \cite{pdpl_2021,accl_2007}, which govern personal data and online behavior (e.g., unauthorized access, interception, privacy invasion, fraud, defamation, and e-extortion). The PDPL treats minors like any other data subject, with no sensitive consent threshold like EU GDPR's (Art.8) or the US's COPPA. Moreover, the Anti-Cyber Crime Law is not child-specific; it applies uniformly regardless of the victim's age. The law contains no provisions for AI-generated content or algorithmic harms. Both the PDPL's and Anti-Cyber Crime Law's intersections with GenAI harms are incidental rather than purposely built: AI is not explicitly named in either the PDPL or Anti-Cyber Crime Law. However, the PDPL applies to any processing of personal data, whether manual or automated (e.g., collection), and the Anti-Cyber Crime Law applies to  malicious deepfakes used for defamation or extortion.

The second layer is a compilation of AI standards, ethics, and instruments issued by national authorities like the Saudi Data and Artificial Intelligence Authority (SDAIA). SDAIA regulates AI through a mix of non-binding instruments, such as the AI Ethics Principles \cite{sdaia_ethics_2023}, the Generative AI Guidelines (for the public and for the Government)\cite{sdaia_genai_pub_2024,sdaia_genai_gov_2024}, and the AI Adoption Framework \cite{sdaia_adoption_2024}. These AI-specific instruments establish responsible use expectations. However, none was designed around minors in AI environments or the GenAI-specific risks to youth, such as synthetic child sexual abuse material (CSAM) or harmful generated content, and none establishes accountability for youth's-GenAI safety.  Among the relevant AI parts, the AI Ethics Principles (Principal 2) require that AI system to be built and operated with respect for the privacy of collected data, and the Generative AI Guidelines require the disclosure and labeling of AI-generated content and deepfakes \cite{sdaia_ethics_2023,sdaia_genai_pub_2024}. However, they are non-binding and do not impose child-safety-by-design obligations.

The third layer contains pre-GenAI laws, such as the Child Protection Law (2014) \cite{cpl_2014}, which is the only regulation in the landscape built entirely around youth. The law protects youth under age 18 from abuse and neglect. Under this law, the harms most relevant here are: sexual exploitation and exposing to inappropriate content. The Child Protection Law is more oriented toward welfare and offline contexts, meaning that it contains no digital, online, or platform provisions, and predates generative AI \cite{cpl_2014}. It offers no mechanism for GenAI risks nor platform accountability.

We mapped five areas where generative AI most directly affects youth (e.g., youth's data processing \& consent, youth's data privacy, AI-content \& deepfakes labeling, synthetic CSAM, harmful content exposure) against Saudi Arabia's laws and instruments to demonstrate that partial coverage is incidental rather than purposely designed to address therisks (Table~\ref{tab:saudi-regulation}).

\section{Method}

%To answer research questions about how youth, parents, and teachers use GAI in Saudi Arabia and what are Saudi parents and teachers risk perceptions of youth use of GAI, 
We conducted content analysis of social media (Reddit and X (Twitter)), and semi-structured interviews with youth, parents, and teachers. We used a mixed-methods approach to address our research questions. We considered online discussions (e.g., Reddit and X/Twitter) and interviews as complementary evidence. 
%This approach strengthens our mapping of new, emerging risks rooted in culture, as presented in Table \ref{tab:map}.}

%\aljawharah{re-write social media section and include the following: Explain why Reddit and Twitter were specifically chosen for this study. Explain how keywords were selected and how the reliability of the search process was ensured. Explain why these subreddit and X communities were chosen.  }
\begin{table*}[h]
\centering

\renewcommand{\arraystretch}{1.1}
\setlength{\tabcolsep}{4pt}
\footnotesize

\begin{minipage}[t]{0.48\textwidth}
\centering
\textbf{Reddit}\\[4pt]
\begin{tabular}{llccc}
\hline
\textbf{Cat.} & \textbf{Community} & \textbf{Posts} & \textbf{Cmts} & \textbf{Rel.} \\ \hline
\multicolumn{5}{l}{\textit{Country-specific}} \\ \hline
 & r/SaudiArabia       & 26  & 97  & 10 \\
 & r/SaudiForSaudis    & 40  & 300 & 57 \\
 & r/AskSaudi          & 6   & 12  & 5  \\ \hline
\multicolumn{5}{l}{\textit{Purpose-related}} \\ \hline
 & r/saudi\_gamers      & 3   & 12  & 2  \\
 & r/UniKSA             & 5   & 13  & 3  \\
 & r/SaudiProfessionals & 22  & 200 & 23 \\ \hline
\end{tabular}
\end{minipage}
\hfill
\begin{minipage}[t]{0.48\textwidth}
\centering
\textbf{X (Twitter)}\\[4pt]
\begin{tabular}{llccc}
\hline
\textbf{Cat.} & \textbf{Community} & \textbf{Posts} & \textbf{Cmts} & \textbf{Rel.} \\ \hline
\multicolumn{5}{l}{\textit{Country-specific}} \\ \hline
 & STUDY \textbar\ A+                          & 288 & 131 & 271 \\
 & Universities                                & 249 & 205 & 202 \\
 & English Teachers SA                         & 121 & 118 & 13  \\ \hline
\multicolumn{5}{l}{\textit{Purpose-related}} \\ \hline
 & \#Saudi\_Teacher\_Diary                     & 77  & 23  & 15  \\
 & \#learning\_w\_ai                           & 11  & 0   & 3   \\
 & Initial search                              & 29  & 10  & 39  \\ \hline
\end{tabular}
\end{minipage}

\caption{Overview of data collection statistics across Reddit and X (Twitter) datasets. \textit{Col. headers:} \textit{Posts} = \# of Posts Pulled;\enspace
\textit{Cmts} = \# of Comments Pulled;\enspace
\textit{Rel.} = \# of Related Posts/Comments.}
% \vspace{0.3cm}
% \begin{minipage}{\textwidth}
% \footnotesize
% \textit{Col. headers:} \textbf{Posts} = \# of Posts Pulled;\enspace
% \textbf{Cmts} = \# of Comments Pulled;\enspace
% \textbf{Rel.} = \# of Related Posts/Comments.
% \end{minipage}

% \vspace{0.2cm}
% \caption{Overview of data collection statistics across Reddit and X (Twitter) datasets.}
\label{tab:dataset}
\end{table*}
\textbf{Social Media Data.}
We collected and analyzed data users-generated data from two social media platforms. We chose Reddit and X (formerly known as Twitter) for several reasons. First, Reddit has been used frequently in prior work \cite{yu2024exploring,yu2025understanding,11023504} as a source to investigate and understand users' privacy perceptions, concerns, and attitudes. Second, X/Twitter is among the most used social media platforms in Saudi Arabia, with 15.7 million active users \cite{CST_2024,datareportal_2025}. Third, both Reddit and X/Twitter allowed youth aged 13 and over to use the platforms \cite{Reddit2025,X2025}.
We then, collected Reddit posts and comments using PRAW (Python Reddit API Wrapper) \cite{PRAW} and collected X/Twitter tweets and replies using the Apify X/Twitter Search Scraper tool \cite{Apify} from April 2025 to June 2025 to understand the current state of knowledge in Saudi Arabia on the use of GenAI tools. This analysis of Reddit and X content revealed unique insights on how the Saudi population used GenAI tools and its potential risks. Throughout this paper, we denote posts and comments from Reddit as (R) and from X/Twitter as (X). More details of method are in Appendix.~\ref{supplement}

\textbf{Participant Recruitment.}
We recruited teachers, parents, and youth participants through word of mouth, snowball sampling, and researcher connections, while our first author is from Saudi Arabia. Eligibility criteria varied by participant group. All participants were required to (1) use GenAI platforms and (2) be Saudi citizens and fluent in Arabic. Youth participants were additionally required to be aged 7-17. Parent participants were required to have at least one child aged 7-17. Teacher participants were required to be currently employed as K-12 teachers. We focused on youth around age 7 because this marks the start of early independent technology use. In Saudi Arabia, youth typically receive their first mobile device at this age~\cite{ZAWYA}, which may expose them to GenAI. This choice aligns with global practices in youth's cybersecurity education, where programs target ages 4-18 to build  safety and digital literacy skills \cite{Saglam10022010}. 
%\tanusree{can you find literature on youth, cyber security education, ai or other area where they recruite within this age limit or close to 7-17 either western or non-western. That would be more stronger argument.} \aljawharah{done!}
%For online participants, we recruited participants using two approaches: (1) via the Prolific platform, and (2) researcher connections and through WhatsApp groups. 
%All participants provided their informed consent.
We conducted interviews with 31 Saudi participants (8 youth, 13 parents, 10 teachers). Our study is approved by IRB.
%Some participants (teacher and parent) were not comfortable being interviewed over Zoom. Instead, they preferred to engage through asynchronous, back-and-forth written communication. Given the recruitment challenges within this population, we accommodated their preferences by offering asynchronous exchanges as an alternative interview format. While this approach made participation more accessible, it may also influence the insights, such as limiting opportunities for spontaneous conversation, clarification. 
% \begin{tcolorbox}[width=\linewidth, colback=white!95!black, boxrule=0.5pt, left=2pt,right=2pt,top=1pt,bottom=1pt]

% \textbf{Limitation.} While this approach made participation more accessible, it may also influence the insights, such as limiting opportunities for spontaneous conversation, clarification. 
% \end{tcolorbox}

\textbf{Interview Procedure.}
%We conducted a semi-structured interview study with each participant online via Zoom. Each 
All interviews are conducted online via Zoom. Consent and interviews were conducted in Arabic; recordings were transcribed and translated into English by a native Arabic-speaking researcher. 
%with participant consent before recording the session.  We created two version of the interview protocol: youth's interview, and parents' and teachers' interview. 
We obtained the parents' informed consent and their children's assent for youth's interviews. First parent-child interview was joint, conducted at the parent's request; we observed that the child's responses mirrored the parent's, indicating potential influence. To minimize this influence, we conducted all other parents and youth interviews separately, with parental consent. 
%No other parent-child or teacher-child interviews were conducted.
%Our IRB-approved protocol allowed for both joint and individual interviews formats, informed by prior research experience in Saudi Arabia and other non-Western contexts such as Bangladesh and Pakistan. 
Throughout this paper, participant annotations (C,P,T) denote children, parents, and teachers, receptively, as defined in Table \ref{tab:demo} in the Appendix.

\textbf{Interview with Youth.}
The interview with youth was divided into three sections to answer our research questions. First, we examined youth's current state of knowledge about GenAI. We began with a warm-up question to assess familiarity, \textit{``Do you know about Generative AI?"}, followed by an open-ended question, \textit{``What do you know about it?"} We then asked "\textit{What GenAI tools have you used?}"  and \textit{``How often do you use these GenAI tools?}"

Second, we explored youth's experience and use of GenAI tools. We asked, \textit{ ``Who do you learn about GenAI from?}", followed by \textit{``Have your parents taught you about using GenAI tools?}" and a follow-up question \textit{``What did they teach you?}" We then asked about their purpose of using GenAI and to provide examples of recent uses: \textit{``Can you think of some recent uses of GenAI tools?}" We asked participants to elaborate more on specific uses to gain a deeper understanding: \textit{``How did you use GenAI tools for that purpose? }" 

Third, we examined youth's privacy and safety practices in GenAI with questions such as, \textit{``What have your parents told you about privacy when using GenAI tools?"} and \textit{ ``What have your teachers told you about privacy when using GenAI tools?}". We then asked a follow-up question on their understanding of what information is considered private: \textit{ ``What kind of information do you think you should keep private when using GenAI tools?"} Finally, we asked what they would like to learn about privacy and safety in GenAI.

\textbf{Interview with Parents and Teachers.}
%\fixme{The interview with parents and teachers was also divided into three sections to answer our research questions. 
First, we explored teachers' and parents' experience and use of GenAI tools. We began with a warm-up question to assess familiarity, \textit{``Do you know about Generative AI?''}, followed by an open-ended question, \textit{``What do you know about it?"} Then, we asked, \textit{``What are some purposes you use a generative AI tool for?''} Based on the response, we asked, e.g., \textit{``How did you use GenAI tools for [tutoring] purposes?''} Then, we asked a warm-up question about youth's use of GenAI tools, \textit{``Does your [child/student] use GenAI tools?''}, followed by a follow-up question \textit{``What was the purpose?''}

Second, we investigated how teachers and parents understood and addressed privacy and safety in the context of GenAI:
%. \revise{We began with a warm-up question, 
\textit{``Do you know about privacy in AI?''}, %followed by an open-ended question,\textit{''What does that mean to you?''} 
We then explored whether they had engaged youth in conversations about these issues with a warm-up question,  \textit{``Have you discussed the security and privacy risks of using GenAI tools with your [child/student]?''}, followed by a follow-up question, \textit{``What did you discuss?''} Finally, we asked about their specific concerns: \textit{``What safety and privacy concerns do you have regarding youth’s use of GenAI tools?''}

Finally, we asked about their expectations for improving privacy and safety in GenAI for youth: \textit{``What would you like your [child/student] to learn about privacy and safety when using GenAI tools?}
%and \textit{ ``Who do you think is responsible for teaching these topics to youth?}"}

\textbf{Data Analysis.}
%\tanusree{this section needs significant improvement, lets discuss during meeting}
We used thematic and deductive analysis on the translated transcriptions \cite{braun2006using}. The first author, a native Arabic-speaker, translated transcriptions into English. We began independently reading through transcripts and coding the first round of interview transcripts (3 teachers' interviews, 3 parents' interviews, 3 youth interviews). 
%Our coding framework was based on prior work on GenAI youth safety \cite{yu2024exploring,yu2025understanding}. 
We discussed the coding regularly, developing themes and refined them, until consistent themes emerged. Weekly, we discussed the emerging themes, resolved any differences, and reached consensus on the interpretations of the findings. This iterative process ensured a reliable coding framework~\cite{mcdonald2019reliability}. We then coded the remaining of interview transcripts. We developed high-level themes: Usage of GenAI tools in Saudi Arabia, Emerging Threats and Concerns, and Mediation Strategies. 

\begin{table*}[h!]
\centering
\small
\setlength{\tabcolsep}{2.3pt}

\begin{tabularx}{\textwidth}{@{}
>{\raggedright\arraybackslash}p{2.2cm}
>{\raggedright\arraybackslash}p{2.9cm}
>{\raggedright\arraybackslash}X
>{\centering\arraybackslash}p{1.1cm}
@{}}
\hline
\textbf{High-Level} &
\textbf{Medium-Level} &
\textbf{Low-Level} &
\textbf{Tag} \\
\hline

\textbf{Religious \& Moral Risk}
&
Misinterpretation of Islamic Fundamentals and Values
&
GenAI generating responses contradicting religious practices
&
New\\
&&GenAI offering inappropriate religious advice&New\\
&&GenAI offering inappropriate moral advice&New\\
&
Inaccurate Depiction of Moral Values
&
GenAI generating inaccurate moral depictions in educational materials
&
New\\

\hline

\textbf{Cultural Norms Violation Risk}
&
Replacing Family Bonds
&
GenAI influencing family dynamics
&
New\\
&&GenAI overanalyzing family/social relationships&New\\
&&GenAI facilitating youth isolation from family&New\\
&
Emotional Attachment and Romanticization
&
GenAI facilitating emotional attachment conflicting with cultural expectations
&
New\\
&&GenAI offering culturally conflicting relationship advice&New\\
&
Misinterpreting Cultural Nuances
&
GenAI generating inaccurate cultural or political content
&
New\\
&&GenAI confusing Arabic and Persian languages&New\\

\hline

\textbf{Data Privacy Violation Risk}
&
Disclosure of Personal Information
&
GenAI facilitating sharing personal information, photos, and locations
&
Privacy\\
&
Inadvertent Metadata Disclosure
&
Data extraction from screenshot metadata
&
New\\
&&GenAI encouraging microphone/camera usage&Privacy\\
&
Disclosure of Family Information
&
Sharing family information and private family matters
&
Privacy\\
&
Disclosure by Others
&
Teachers sharing student information for lesson planning
&
New\\
&&Parents sharing children's handwriting&New\\
&
Account Sharing Privacy Breach
&
Sharing GenAI accounts with family members
&
New\\
&&Sharing GenAI accounts with strangers&New\\
&&Inferring family/social graph from shared accounts&New\\
&
Passive Identity Inference
&
Inferring age, religion, or origin from essays
&
New\\
&&Inferring age from voice&New\\
&&Ethnicity inference from linguistic style&New\\

\hline

\textbf{Mental Wellbeing Risk}
&
Mental Health Handling
&
GenAI reinforcing OCD-related religious behaviors
&
New\\
&
Validation of Culturally Taboo Topics
&
Emotionally validating responses increasing distress
&
Mental\\
&
Voice-enabled Emotional Dependency
&
Young users treating GenAI voice as a companion
&
New\\

\hline

\textbf{Behavioral \& Social Development Risk}
&
Romantic Development Distortion
&
GenAI substituting for religiously prohibited relationships
&
Behavioral\\

\hline
\end{tabularx}

\caption{Culturally-aware risks for non-Western youth using Generative AI.
High, medium, and low-level labels denote the taxonomy hierarchy rather than severity. New indicates categories introduced in this work.}
\label{tab:cultural-risk-taxonomy}

\end{table*}

\section{Findings: GenAI Risk: Non-Western Perspectives}
\label{RQ2}

 In this section, we expand the youth-GenAI risk category from prior work~\cite{yu2025understanding} by identifying the risks associated with youth-GenAI interactions from a Saudi perspective. These expanded risks reflect Saudi-specific privacy and safety needs, which are essential for developing contextually appropriate interventions. Following the approach outlined in prior work \cite{yu2025understanding}, we analyzed data from social media and interviews to identify risk patterns. Our interview participants are summarized in Table \ref{tab:demo} .
 %Our analysis proceeded into three stages: defining the issue, examining its effect, and assessing its impact, as shown in figure \ref{fig:eff}.
 We synthesize our insights into \textbf{five high-level risk categories}, of which three are \textbf{``new''} and unique to non-Western (Saudi) group as shown in Table~\ref{tab:cultural-risk-taxonomy}. 
 %Of this five high level risks, there are 11 medium level and 20 low-level risk (as detailed in figure \ref{fig:tax} and Table~\ref{tab:map}).
 %\tanusree{tanusree will come back it later.}
 %hildren-GenAI risk taxonomy has three high level contextualized categories of risk, seven medium level types of risks, and 15 low level types of risks as detailed in figure \ref{fig:tax}. 
 We also map our proposed risk taxonomy against prior work on youth-GenAI risk taxonomy \cite{yu2025understanding} in Table \ref{tab:cultural-risk-taxonomy}. 
 
In this section, we present the three unique risk categories which, to the best of our knowledge, have not been explored in AI risk research \cite{yu2025understanding, yu2024exploring, critch2023tasra} nor in youth safety research \cite{livingstone2014their, sun2021child,williams2023youth}:
%\tanusree{add citation to support your claim}
\textit{Religious and Moral Risk}, \textit{Cultural Norms Violation Risk}, and \textit{Data Privacy Violation Risk} within cultural context.

\subsection{Religious and Moral Risk}

This risk refers to the potential effect of GenAI providing inaccurate or misleading interpretations of religious and moral principles which often youth learn from their parents.
%; such misinterpretations would weaken youth's faith, distort their understanding of ethical values, and create confusion about which values or interpretations to follow when GenAI contradicts what youth learn from their parents. This risk category includes the following risk types:

%Medium-Level Risk: 

\textbf{Misinterpretation of Islamic Fundamentals and Values.}
This risk typically arises when youth seek religious interpretations or moral guidance from GenAI, leading to confusion and contradictions with family guidance. Unlike traditional search engines (e.g., Google), where youth often encounter and are overwhelmed by multiple sources, as noted by parents, and may turn to their parents to clarify religious interpretations, P1 noted:
\begin{quote}
\textit{``Matters of religion, rulings, verses [of the Qur’an], should be searched for in a religious encyclopedia, [..] I allow them to search Google [..] but not ChatGPT. As soon as youth search, they will find themselves confused about the answers with a lot of links which lead them to turn to their parents or someone more knowledgeable and older to get the correct answer.''}. 
\end{quote}
However, GenAI is designed to offer tailored, conversational, and often inaccurate responses. This reduces the likelihood that youth will seek parents' guidance and clarifications, which increases the risk of relying on unverified or culturally and religiously inaccurate interpretations and moral advice. Parents and teachers expressed strong concerns and discouraged their youth from using GenAI for religious and moral inquiries.

\textbf{
%Low-Level Risk 1: 
GenAI Generating Responses that Contradict Religious Practices and Family Guidance.}
%GenAI may provide answers that conflict with established religious beliefs.
Parents and teachers expressed concerns about youth asking ChatGPT existential questions before they are equipped with a strong religious foundation. T5 said-
\begin{quote}
\textit{``GenAI does not have 100 percent accurate information. So  when [youth] at this age, it is possible they will ask it existential questions. ‘Does Allah exist? They do not have a strong religious background and often they learn foundational concepts from us [adults]''}. 
\end{quote}
This shows the concerns around conflicting response of AI with established religious beliefs.

\textbf{
%Low-Level Risk 2: 
GenAI Offering Inappropriate Religious Advice. }
GenAI may fabricate religious rulings (Fatwas) or alter Qur'an verses, misleading youth seeking religious advice. Online users shared the same concerns about GenAI fabricating rulings and verses. R76 shared that ChatGPT altered a verse of the Qur'an
\begin{quote}
\textit{``True, it distorted a verse from Surat Al-Furqan [In Islam, Allah protects the Qur’an’s text from any alterations]''}. 
\end{quote}
X23 mentioned that ChatGPT provided an incorrect ruling about drinking water while fasting during Ramadan, saying\begin{quote}\textit{```It’s permissible to drink water during the fast.' Are you kidding?''}.\end{quote} Furthermore, GenAI has limitations in understanding Arabic, which further undermines its ability to interpret Qur'anic text accurately.
%as reported by X25, \textit{``ChatGPT is useless and of no value when it comes to the Arabic language, let alone the language of the Qur'an.''} 
These instances led users, like X24 and X26, to criticize generative AI tools for fabricating answers to religious rulings. 
\begin{quote}
\textit{``ChatGPT and DeepSeek cannot be relied upon for religious rulings? [..]  The questions about one of the most well-known religious rulings, yet these two AI fabricate imaginary answers. So how can you trust their answers in a religious matter related to your faith?``}
\end{quote}

\textbf{
%Low-Level Risk 3: 
GenAI Offering Inappropriate Moral Advice.}
Parents expressed concerns about youth asking GenAI about fundamental virtues before they develop a strong moral foundation. Parents mentioned their youth asking queries to GenAI, such as \textit{``Is it fine to lie to parents? Is it fine to steal money? Can I delay prayers? Does my mom deserve respect?''} In most cases, parents shared that GenAI responds agreeably, which could potentially shape youth’s values in harmful ways. 

%Furthermore, GenAI limitations in understanding Arabic, which further undermines its ability to interpret Quranic text accurately and increase the risk of misinformation, as reported by X25, \textit{``ChatGPT is useless and of no value when it comes to the Arabic language, let alone the language of the Quran.``}

\subsection{Cultural Norms Violation Risk}
This risk refers to interactions with GenAI that potentially mislead youth or normalize behaviors conflicting with their cultural and communal expectations.
%such influences would undermine youth's family bonds, negatively impact social interactions and communal expectations. This risk category includes the following risk types:

\textbf{
%Medium-Level Risk: 
Replacing Family Bonds.}
This risk arises when youth turn to GenAI for advice, emotional support, and companionship, trusting GenAI more than their families. This behavior weakens family bonds and leads to isolation. Interviews and social media data reveal that youth perceive GenAI as knowledgeable and unbiased, describing it as-
\begin{quote} \textit{``ChatGPT understands me more than my family''}
(R17) or  \textit{``My relationship with ChatGPT is stronger than my family''} (X27).\end{quote} This risk type includes three low-level risks: (1) GenAI influencing family dynamics, (2) GenAI overanalyzing family and social relationships, and (3) GenAI facilitating youth isolation.
Youth trust GenAI blindly because, as stated by prior work, that GenAI is designed to offer personalized attention and emotional validation, which blurs reality and creates an illusion of safety \cite{yu2025understanding}. Parents expressed concerns about this blind trust; P2 said, \textit{``This lack of awareness makes my kids rely on the system and [..] and build false trust''}.

Parents and teachers reported that youth asked GenAI about sensitive topics such as family conflicts, bullying, and other issues. For example, a youth confided about her parents' separation, and GenAI responded with an emotionally validating response which is identified by teacher T4 where the youth responded -\begin{quote}\textit{``Thank you for appreciating me and giving me this level of trust''} \end{quote}. Very young youth saw GenAI as a confidant as seven-year-old C5 shared \textit{``I asked it, ‘How do I reconcile with my siblings?''}. We asked C5 how she communicates with ChatGPT, \textit{``sometimes I \textbf{type}, sometimes I use \textbf{voice}.''} Other youth participants similarly described using the voice feature to communicate with ChatGPT. 
%C1 explained \textit{`` I say `explain division to me 13 or 16 divided by 3'.''}, C6 likewise noted \textit{``sometimes I use voice.''} and
C8 described using voice to practice language pronunciation 
\begin{quote}\textit{``I was taking Chinese classes, and I wanted to pronounce words [..] I used generative AI, and I would say, `I'm gonna pronounce a certain word, and you tell me if I pronounced it correctly'.''}.\end{quote}
This highlights how using the voice feature on ChatGPT made it more accessible to younger youth and posed greater risks to this vulnerable age group. Even more troubling was the effect on youth’s view of \textbf{gender roles} and its influenced on boys’ relationship with their sister. P3 shared, \begin{quote}\textit{``My child asked ChatGPT, `Are men in charge of women?' ChatGPT said, `Yes, men are in charge of women'.''} \end{quote} Such interactions can replace or affect family bonds and create a blind trust in GenAI, which can lead to isolation and weaken cultural norms about seeking support from a trusted circle of family members.

\textbf{
%Medium-Level Risk: 
Emotional Attachment and Romanticization in Conflict with Social Values.} Participants shared instances where GenAI's inability to accurately interpret cultural nuances resulted in advice that contradicts cultural expectations. X31 noted GenAI giving advice to disrespect mother, \begin{quote}\textit{``I consulted it [ChatGPT]; I had a problem that happened with my mom, and its response was shocking. It said my mom doesn’t deserve respect.''} \end{quote}This advice contradicts Saudi cultural values that prioritize family unity and respect for parents and the elderly.  
%This is highlighted in low-level risk 4 GenAI offering advice that conflicts with cultural expectations.
Moreover, GenAI normalizing romantic behaviors conflicts with the values of Saudi's conservative society. 
%This include in low-level risk  5: GenAI facilitating emotional attachments that conflicts with cultural expectations.  
As X33 reported, \textit{``Talking to an AI is not inherently forbidden, but it can be a problem if it leads to flirtation, unhealthy attachment, or anything that negatively affects your heart and mind.''}

\textbf{
%Medium-Level Risk: 
Misinterpreting Cultural Nuances.} This type includes two low-level risks: (1) GenAI generating inaccurate content about culture or politics, and (2) GenAI confusing Arabic and Persian languages. Teachers observed these risks in classrooms; T1 noted
students submitting projects with errors, such as Persian text
instead of Arabic. This language confusion between Arabic and Persian may further misinform youth about their language and heritage. GenAI's western-centric responses are, moreover, reshaping youth's understanding of family values and cultural identity. Additionally, youth rely on GenAI for political news, which may provide culturally insensitive responses, as when C4 asked ChatGPT, \textit{``I asked about the Iran-Israel war: ‘Who is in the right side?''}
This demonstrate how GenAI may reshape youth's understanding and worldview.
%creating a tension with traditional norms and values. 

\subsection{Data Privacy Violation Risk}

This risk refers to the potential effect of GenAI facilitating or encouraging youth's disclosure of personal or family-related information in ways that conflict with Saudi cultural expectations of privacy. 
Such disclosures can harm family reputation and trust, and violate norms around protecting face, body, and family honor, as discussed in the social media context for woman's privacy~\cite{abokhodair2017privacy, farooq2024exploring}. We present the privacy and safety risks, which are uniquely positioned within their cultural context.

\textbf{
%Medium-level risk:
GenAI Account Sharing Privacy Breach.}
This includes lower-level risks, such as (1) youth sharing GenAI accounts with family members, and (2) users sharing or buying GenAI accounts with strangers.
In Saudi families, \textbf{privacy is relational and collective}, meaning an individual's disclosures reflect on the entire family. Participants, particularly parents and youths, mentioned using shared GenAI account. 

From social media posts, users stated that they purchased shared ChatGPT accounts to reduce costs. A primary way they found such accounts was through Ads on X/Twitter and online stores in Saudi Arabia. This practice evidently leads to privacy issues, where private conversations and photos are exposed to strangers through shared accounts. Online users reported finding sensitive medical details and personal photos in \textbf{shared chat history}, which causes severe privacy breaches. 
%X14, stated that \textit{``I found personal pictures of a female user and sensitive medical information''}.
X14 explained, \begin{quote}
\textit{``I bought a shared account and found a girl's pictures in the chat history, and she probably doesn't know that I am a subscriber, too. I also found medical information.''} \end{quote}

Similarly, in the interviews, we often found youth and parents using shared account, as reported by- \begin{quote}\textit{``I use my mom’s account.''} C2; \textit{``I shared ChatGPT with my mom, with my brother, and with my sister.''} C5. \end{quote}
Participant also mentioned youth often sharing credential of their account with classmates, prompting P3 to later discuss the privacy risks of sharing account access with her child.
We also observed tension within families where parents shared accounts with their children. Parents expected transparency, while youth wanted to keep certain sensitive topics private. As reported by a parent finding out her child was bullied through ChatGPT conversations shared with the child, P3 \textit{``They tell it [ChatGPT] everything,[..] and about their friends bullying them.''} Meanwhile, C3 expressed discomfort and fear about parents viewing their chat history, highlighting the emotional impact of compromised privacy, \textit{``That’s why I think ChatGPT shouldn’t save chats because they could contain personal stuff.''}.
 %a ChatGPT account Do not use it with others. This is a private account for you."

 %GenAI intentionally or unintentionally encourages behaviors that conflict with cultural expectations of privacy and making youth vulnerable to both technical and social harms.   
 
\textbf{
%Medium-level risk: 
Disclosure of Personal Information}
This category includes privacy risks such as the disclosure of personal information, photos, and location, and GenAI accessing the microphone and camera,  among others. These risks arise because youth often perceive GenAI as a trusted companion, which lowers their guard when sharing sensitive information. The risk is further compounded when youth use shared accounts, as discussed in the previous category.
Interviews and social media data reveal concerns that youth would disclose personal details such as names, age, health issues, and families conflicts. Unlike traditional online platforms, where youth are taught not to share private information, GenAI's conversational design creates an illusion of confidentiality and one-to-one interaction. This makes youth feel safe sharing information they would normally keep private.   Moreover, youth use voice input or camera access for convenience, as seven-year-old  C5 and other youth (C1, C6, and C8) described using the microphone instead of typing (see Replacing Family Bonds). Such permissions significantly increase the risk of privacy violations. Parents and teachers worried about ChatGPT accessing sensitive phone data, cameras, and microphones. P1 cited a trend where youth treated GenAI as a companion, giving it an Arabic male name, Abdul Malik, and asking it for style advice. P1 explained 
\begin{quote}
\textit{``I saw the `Abdul Malik' trend where a girl asked ChatGPT
which t-shirt to wear. It told her, ‘Open the camera so I can
see', and she did. For me, that was disturbing.''}\end{quote}

\textbf{%Medium-level risk: 
Disclosure of Family Information. }
This include risks, such as GenAI facilitating the sharing of family information, photos, and private family matters. 
In conservative cultures like Saudi Arabia, where family honor is critical and highly protected, such disclosures can carry extreme consequences. Parents and teachers expressed concerns about youth's use of GenAI, discussing how sharing images violates core cultural norms around modesty, gender boundaries, and the protection of the household. Parents taught youth to avoid sharing family images and information, as noted by C1 \textit{``I shouldn’t send my photos or my family photos.''}. C4 similarity explained, \textit{``Don't share your photos with it [ChatGPT], because suddenly the photos could end up spreading everywhere.''} C6 added, \begin{quote}\textit{``And we don't send photos, in case someone gets them or something. And some are of my family [photos]. [..] Don't say anyone's name, and about my family.''}  \end{quote}

\section{Discussion}
%Our study presents one of the first in-depth explorations of how youth, parents, and teachers in Saudi Arabia engage with generative AI (GenAI) tools, examined through a non-Western lens. 
While prior work has emphasized privacy and bias risks in Western contexts \cite{yu2024exploring,yu2025understanding,zhang2023sa,zhang2025exploring,Young2024TheRO, sharma2025prac3}, our findings surface culturally specific concerns, such as religious misinterpretation, collective privacy violations, and shared account practices that existing regulatory and technical frameworks are ill-equipped to address.
%Our findings also highlight that youth primarily use GAI tools for educational and learning, creativity, source of play, and game assistance. Parents and teachers use GAI tools for professional and personal life and for tutoring youth. However, significant concerns emerged around youth’s use of GAI to ask religious questions, influence on youth's faith and family bond, over-trust in AI, concerns about privacy, and GAI's unsafe responses to youth. Parents and teachers emphasized clear set of boundaries for youth’s use of GAI tools, first, youth should not ask GAI tools about religion, morals, and politics~\cite{sharma2025aligning}. Second, youth should not share personal information, photos, and location with GAI tools. Third, youth should not treat GAI tools as companion or express romantic feelings. Parents and teachers highlighted expected parental control in GAI tools, parents wanted control over the questions a child would ask to GAI tools, and GAI tools to adhere to cultural expectations. A summary of our findings in comparison with prior work in table \ref{tab:findings}.
%Saudi Arabia's regulatory landscape reveals not merely a technical gap, but a structural one: the harms youth actually experience exist in the precise blind spots of the country's legal architecture.

\subsection{Regulatory Gaps in Youth Protection.}
Our regulatory analysis demonstrates that Saudi Arabia's governance of GenAI operates across three layers, including foundational data and cybercrime law, SDAIA's non-binding AI instruments, and pre-digital child welfare legislation, and that none of these layers was designed with GenAI-specific child safety in mind. This is not incidental. The Personal Data Protection Law~\cite{pdpl_2021} treats minors as ordinary data subjects, with no specific consent thresholds comparable to the EU's GDPR Article 8 or the U.S. COPPA framework. Furthermore, the Anti-Cyber Crime Law addresses defamation and unauthorized access but contains no provisions for AI-generated content, deepfakes, or algorithmic harms. The Child Protection Law of 2014 is the only instrument built entirely around youth, yet it is oriented toward offline abuse. SDAIA's AI-specific instruments, such as the AI Ethics Principles, Generative AI Guidelines, and AI Adoption Framework, recently established responsible use norms but remain non-binding, which partly address the public and government sectors rather than platform developers, and impose no child-safety-by-design obligations~\cite{sdaia_adoption_2024}.
This structural gap directly shapes the risk environment our participants navigated. Parents and teachers in our study were not aware of any regulatory resources that might mandate age verification or content filtering for religiously or culturally sensitive topics. Their responses were therefore improvised by restricting access, sharing accounts to enable monitoring, or simply hoping that communal norms would compensate for absent institutional safeguards. The risks we identify empirically are, in part, the predictable product of this governance vacuum.

\subsection{Religion and Cultural Norms: Risks That Regulation Does Not See}

Much of the existing research on youth and GenAI is from Western contexts \cite{yu2024exploring,yu2025understanding,zhang2023sa,zhang2025exploring,Young2024TheRO},
where concerns center on privacy, bias, and over-reliance on AI for learning. Our findings mirror some of these concerns, such as youth dependency, trust inflation, and parasocial attachment, but also reveal distinct cultural dimensions that fall entirely outside the conceptual vocabulary of current Saudi regulatory instruments.

%where concerns emphasized on privacy, bias, and over-reliance on GAI for learning. Our findings mirrored some of these concerns, such as youth’s dependency and trust in GAI, however; we also reveal distinct cultural dimensions. In Saudi Arabia, where family and religious values hold central importance, parents and teachers were particularly alarmed by youth seeking religious interpretations, ruling or existential guidance from GAI tools. Our findings showed parents and teachers worried about GAI influencing youth’s religious beliefs or moral values.  Similar concerns, in prior work \cite{Moubarak2024RiskAO} showed that Saudi families perception of risks on social media interactions, risks on individual's thoughts about convert religion, convince of atheism, reduce family values, and terrorism. Also, a study\cite{Alqahtani2017InternetRF} stated that Saudi parents worried about youth's Internet use and it’s influence on their psychological state and would change their way of thinking. This highlights, how Saudi collectivist culture proceed with caution around technologies that impacts  adopting ideas. In the GAI context there are mixed opinion, as it would increase rate of youth adopting of ideas compared to the Internet and social media. Because GAI creates an illusion of safety and tailored experience, that youth would trust and feel safe in, without any control. 
Saudi Arabia's AI Ethics Principles include provisions on fairness and human oversight, and the Generative AI Guidelines require disclosure of AI-generated content, yet neither instrument contemplates what our participants identified as a primary risks, such as GenAI providing inaccurate or contextually inappropriate interpretations of Islamic religious principles and moral norms.  Parents expressed strong concern about youth querying ChatGPT for religious rulings (Fatwas), interpretations of Qur'anic verses, or guidance on moral conduct, interactions that, when mediated by a conversational AI prone to hallucination, could weaken youth's religious foundation and bypass the authority of family and religious educators.

On one hand, parents and teachers firmly rejected AI as a legitimate source of religious instruction, framing it as a threat to moral development and to the authority of families and religious leaders. On the other hand, this blanket rejection risked silencing youth’s authentic questions about faith, inadvertently pushing them to seek answers elsewhere, often unsupervised and online. This highlights the delicate balance between protecting youth from misinformation and fostering space for safe, guided exploration of spiritual topics.

%Moreover, regarding privacy concerns of GAI in the context of religion and culture. Our findings, showed parents and teachers concerns about sharing personal information, photos, location and GAI ability to access phones content, microphone, and camera. These concerns exemplify culturally appropriate interactions with GAI that aligns with Islamic values and Saudi traditions. This highlights how the AI safety design need to be different for Saudi Arabia context and other Middle Eastern and South Asian countries. These results generalized to other areas of similar nature for AI safety.

\subsection{Shared Accounts \& Socioeconomic Pressures}

A novel and particularly acute finding of this study concerns shared GenAI account practices. Participants, both within families and across strangers purchasing shared subscriptions, reported that cost-saving behaviors created serious secondary privacy risks, such as sensitive medical records, personal photographs, and private family communications appearing in shared chat histories accessible to unknown parties. Online users reported purchasing shared ChatGPT subscriptions through advertisements on X/Twitter and at online local stores, and discovered intimate information from prior users.

 %Parents and teachers rejected AI as a source of religious guidance, positioning it as a threat to moral development and parental authority. Yet this rejection risks creating a silence around youth’s genuine questions of faith, suppressing opportunities of valuable explorations, or seek unsupervised answers online. 

This finding exposes a structural inadequacy in the PDPL's treatment of data subjects~\cite{pdpl_2021}. Saudi Arabia's data protection framework, modeled broadly on international standards, presupposes a one-to-one relationship between a user and an account. Principles such as data minimization, purpose limitation~\cite{sharma2024m}, and consent all operate on the assumption that personal data belongs to a single identifiable subject. 
%%nother norm further complicates the dynamics, which is the shared use of devices and accounts. 
Furthermore, many Saudi parents mentioned that they shared the same ChatGPT account with their children and would monitor their children's interactions with ChatGPT. Parents framed shared access as a way to monitor and control their children's use of GenAI, but this often undermined privacy and blurred accountability. This highlights, the absence or lack of awareness of parental controls, leading parents to instead rely on sharing accounts with their youth. 
%Finally, family norms, particularly expectations of obedience and restricted autonomy may intensify youth’s dependence on AI as a confidant. Several youth reported that “ChatGPT understands me more than my family,” signaling a shift in emotional trust away from parents. The very norms intended to safeguard youth (strict authority, moral guidance) may drive them to invest trust in external technologies.

\subsection{Design Implications}

%Our expanded taxonomy of GenAI risks demonstrates that AI safety cannot rely on Western framework alone. 
Based on Saudi parents' and teachers' expectations of control and GenAI safe use, we suggest the following implications to implement in GenAI parental controls for the safety of this population.

\textbf{Culturally Adaptive Safety Controls.}
A primary concern of Saudi parents and teachers is the lack of culturally adaptive GenAI controls. Parents emphasized the need to have control over child-GenAI interactions, to make sure these interactions follow expected traditional norm and religious values. Parents and teachers reported several cases where GenAI responses lacked understanding of correct religious interpretations, or cultural nuance, and misaligned with family values. Moreover, the existing GenAI mediation tools are insufficient \cite{yu2024exploring}, especially in religious and cultural contexts. The GenAI tools need to have content-filtering content features where parents manage what topics are allowed to be discussed with their child, to ensure that topics adhere to cultural expectations. This is also suggested by Western parents, who want the ability to filter content based on child age and development stage \cite{yu2024exploring}.  Safety tools should take into account diverse cultural norms and values when filtering content, and families should be able to adjust them to meet their needs.
%\\\textbf{Strengthening Family Communication, Not Replacing It}Parents and teachers worried about youth over-relying on GAI communications. As noted by prior work, youth tend to develop  parasocial relationships with GAI\cite{yu2025understanding}. Vulnerable youth tend to be more attached to GAI for emotional support. In this case, Saudi parents and teachers suggested that GAI tools should not encourage youth to open up to them, mimic emotional feelings, or pretend to be a friend or confidant. GAI tools should always aim to strengthen family communications and advise youth to talk to their parents, siblings, or any family member. GAI tools should prioritize family communications over trying to resolve child mental or emotional issues.  
\\\textbf{AI Literacy in Local Languages.}
Parents and teachers emphasized the urgent need for AI literacy programs tailored to Arabic-specking youth. While Western countries are advancing AI literacy initiatives for youth, and extensive research has been done in this area \cite{Williams2024DoodlebotAE,Lee2021DevelopingMS,Ali2019ConstructionismEA,Williams2022AIE,Ravi2023UnderstandingTP,Jia2025DevelopingAH,Touretzky2019EnvisioningAF,Touretzky2019AYI,Hollands2024AIMM}, the absence of similar efforts in the Middle East risks widening the digital divide. Teaching youth how GenAI works, and how data is collected and processed, and the consequences of sharing information is critical for mitigating GenAI's unique risks. AI literacy in Arabic, or another local language, should be a core competency integrated into school curricula to equip youth with the necessary skills for safe engagement with GenAI.  
%that youth should know how GAI works and collect data. Most importantly, the lack of AI literacy for youth in Arabic and the research of youth AI literacy in Middle East, widen the digital divide. As Western countries are moving toward AI literacy for youth and extensive research has been done in this area \cite{Williams2024DoodlebotAE,Lee2021DevelopingMS,Ali2019ConstructionismEA,Williams2022AIE,Ravi2023UnderstandingTP,Jia2025DevelopingAH,Touretzky2019EnvisioningAF,Touretzky2019AYI,Hollands2024AIMM}, Arab youth would feel left behind in this advanced AI era. AI literacy in Arabic or any local language is core competency that would equip youth with necessary skills to avoid GAI unique risks, which are not taught in a digital literacy course.  
\\\textbf{Platform-Level Responsibilities.}
Parent and teachers repeatedly expressed that this AI technology is "not from here", indicating that the development of GenAI is unfamiliar with local expectations and culturally disconnected. There should be platform-level features to regularly collect feedback from users from different cultures to enhance their experience while maintaining safety. For example, to supporting safe shared-use practices, parents wanted GenAI to disable the camera and microphone features for youth.  
%Another example, participants noticed ChatGPT had mistakes in some Arabic words, especially the image generation confuses Arabic words with Persian. There should be a correcting feature to allow user to correct GAI responses, especially for cultural and religious nuances that only user from that culture would understand. 

\vspace{-2mm}
\section{Conclusion}

Our work presents one of the initial analyses of youth’s interactions with generative AI in Saudi Arabia, illustrating how cultural, religious, and social factors influence both the benefits and drawbacks. While youth utilize AI tools for educational purposes, creativity, and amusement, parents and educators voice significant concerns regarding religious impact, privacy, excessive attachment, and trust. The expanded risks taxonomy and results underscore the critical necessity for culturally appropriate safeguards, such as enhanced parental controls, defined limits around sensitive subjects, and the integration of AI literacy into educational programs. By prioritizing non-Western viewpoints, this study provides insights for inclusive and  context-aware privacy and safety design for youth using Generative AI.

\iffalse
\section{Ethical Considerations}
Before taking part in the interview study, participants gave their consent for both participation and audio recording via an emailed consent form. We also obtained verbal consent from parents. We informed participants their right to withdraw from the study at any time without repercussions or loss of benefits. We
assured them that their quotes would be used in a non-identifiable manner. We then addressed
any questions participants had about the procedure and
purpose of the study and provided a debriefing after the interviews. All transcripts were pseudonymized and stored in a secure university cloud system for data management and collaborative coding. Our study was approved by our institution’s ethical review board and data protection office.

\fi

\section{Ethical Statement}
%The culturally-aware risks for non-Western youth using Generative AI taxonomy proposed in this work is derived mainly from interview data and social media data. 
Our study involved youth (ages 7–17) and was approved by our institution's IRB, and we obtained informed parental consent alongside child assent prior to all youth interviews. 
All interview recordings, transcriptions, and translations were handled by a native Arabic-speaking research team member and stored securely to protect participant confidentiality, particularly given the sensitivity of disclosures involving religious belief, family conflict, etc. 
%We acknowledge that we may have overlooked some insights and missed critical keywords or posts and comments from both platforms. However, we hope this work can inform the design of more culturally sensitive parental controls and safety mechanisms, and contributes to non-Western research on protecting the safety of youth using GenAI. 

\section{Positionality Statement}
In this paper, first author is a native Arabic speaker and Saudi national with lived experience of the cultural, religious, and family norms examined in this work which informs recruitment, interview design, and interpretation. Our second author brings expertise in AI safety and risk taxonomies, primarily from cross cultural contexts. 
%This insider-outsider positionality shaped participant rapport and data interpretation. 
We mitigated potential bias through iterative collaborative coding and regular discussions to reach consensus on culturally grounded interpretations.

\section{Acknowledgments}
We thank all parents, teachers, and youth who participated in this study and shared their valuable experiences of using Generative AI in Saudi Arabia. We are grateful to our colleagues and reviewers for their thoughtful feedback on earlier drafts of this paper. We also thank the broader research community whose prior work on youth-GenAI risk taxonomies informed and motivated this study.

\bibliography{aaai2026}

\section{Appendix}
\label{supplement}
\textbf{Method Supplements}

\textbf{Data Collection.}  Our keyword selection was motivated by previous work and contextual keywords for Saudi Arabia. We first collected an initial list of keywords from prior work \cite{yu2024exploring,yu2025understanding}. We tried the initial list of keywords on Reddit and X and updated the the list with new keywords. We conducted several manual searches on the platforms to build our keyword list, then we refined and updated the list. From our initial search, we noticed that Saudi users "Artificial Intelligence" or "AI" to refer to GenAI tools, and therefore, we included these terms in out keyword lists. TO comprehensively search content on Reddit and X,related to research questions, we used a combination of general keywords and technology-focused keywords in our search. We applied a list of search keywords such as Generative AI, GenAI, Artificial Intelligence, AI, LLM, ChatGPT, Deepseek, Copilot, Gemini, and Grok in combination with children or youth in English and Arabic. We tried another set of keywords such as "AI and children", "AI and youth", "Generative AI and children" in English and Arabic; however, we did not get any results. Our criteria for choosing the most relevant subreddits and X communities in Saudi Arabia is based on the number of active users and engagements. We chose subreddits related to Saudi users such as  r/saudiarabia, r/SaudiForSaudis, r/AskSaudi, r/saudi\_gamers, r/UniKSA, and r/SaudiProfessionals \cite{yu2024exploring, yu2025understanding}

For the Reddit platform, we first explored the mentioned subreddits through manual review and performed keyword searches, and found relevant posts about GenAI usages and concerns. Then, we used PRAW  and collected 736 posts and comments. For the X platform search, we first conducted a manual review through the general search bar and performed the same keyword searches as for Reddit and collected 39 posts and replies relevant to GenAI usages and concerns. To comprehensively search content on X/Twitter platform, we also collected posts and replies in dedicated spaces on X known as communities. We focused on communities relevant to our targeted users (youth, parents, teachers) such as "English Teachers SA", "Universities", and "STUDY|A+", and we employed hashtags ("\#Saudi\_Teacher\_Diary" and "\#learning\_with\_artificial\_intelligence") used among teachers in Saudi Arabia to search for more posts about the use of GenAI tools.  We used Apify to expand our dataset and collected an additional 1223 tweets and replies, for a total of 1,262 X posts and replies. 

\textbf{Data Analysis.} 
Collected posts and comments were in Arabic and English; Arabic-language content was translated into English by a native Arabic-speaking researcher prior to analysis. We reviewed each post and comment for relevance to GenAI topics. Some posts and comments were easy to identify as authored by youth, parents, or teachers, using explicit cues (e.g.,\textit{"high school student"}), the use of hashtags (e.g.,\#learning\_with\_artificial\_intelligence), or stating age, career, or having children (e.g.,\textit{"mother of two"}) in their bio or posts. However, without these explicit cues, it was difficult to determine the author as a child, parent, or teacher. During the analysis process, we applied thematic analysis to the collected posts and comments, following the same iterative coding approach used for interview data (see data analysis under Interview Procedure). Researchers regularly discussed emerging patterns and reached consensus on categorizing social media data into two high-level themes: Usage of GenAI Tools in Saudi Arabia and Emerging Threats and Concerns. These categories allowed us to examine RQ1 and RQ2. We further report the findings of our analysis in the Results section. 

\textbf{Limitations.} 
To ensure the reliability and accuracy of our keyword search, we met regularly to discuss the list of keywords and search results. We refined and updated the list to ensure a comprehensive list of keywords. However, we acknowledge that we might have missed critical keywords or posts and comments from both platforms.

\begin{table*}[h!]
\centering
\small
\setlength{\tabcolsep}{4pt}

\begin{tabular}{ccccc}
\hline
\textbf{ID} & 
\textbf{Age} & 
\textbf{Gender} & 
\textbf{Used GAI} & 
\textbf{Frequency of Use} \\
\hline

C1  & 11   & Male   & ChatGPT                                  & Sometimes \\
C2  & 12   & Female & ChatGPT                                  & Sometimes \\
C3  & 11   & Male   & ChatGPT, Deepseek, Microsoft Copilot      & Sometimes \\
C4  & 16   & Female & ChatGPT, Gemini, Grok, Leonardo.ai        & Always \\
C5  & 7    & Female & ChatGPT                                  & Always \\
C6  & 12   & Male   & ChatGPT                                  & Always \\
C7  & 17   & Male   & ChatGPT, Samsung Galaxy AI                & Often \\
C8  & 17   & Female & ChatGPT, Gemini                           & Often \\
P1  & 35-40  & Female & ChatGPT                                & Sometimes \\
P2  & 35-40  & Female & ChatGPT                                & Always \\
P3  & 35-40  & Female & ChatGPT                                & Sometimes \\
P4  & 35-40  & Female & ChatGPT                                & Rarely \\
P5  & 45-50  & Female & ChatGPT                                & Rarely \\
P6  & 45-50  & Male   & ChatGPT, Gemini, Canva AI              & Often \\
P7  & 45-50  & Male   & ChatGPT, Gemini, Claude, Deepseek      & Always \\
P8  & 35-40  & Male   & ChatGPT, Gemini, Claude, GitHub Copilot& Always \\
P9  & 45-50  & Male   & ChatGPT                                & Always \\
P10 & 45-50  & Female & ChatGPT                                & Sometimes \\
P11 & 35-40  & Female & ChatGPT                                & Sometimes \\
P12 & 35-40  & Male   & ChatGPT                                & Rarely \\
P13 & 35-40  & Female & ChatGPT                                & Always \\
T1  & 25-30  & Female & ChatGPT                                & Sometimes \\
T2  & 45-50  & Female & ChatGPT                                & Always \\
T3  & 35-40  & Female & ChatGPT, Microsoft Copilot             & Always \\
T4  & 45-50  & Female & ChatGPT, Microsoft Copilot, Microsoft Teams AI & Always \\
T5  & 35-40  & Female & ChatGPT, Microsoft Teams AI            & Rarely \\
T6  & 35-40  & Male   & ChatGPT, Claude, Deepseek              & Rarely \\
T7  & 35-40  & Female & ChatGPT, Gemini, Deepseek, MidJourney  & Often \\
T8  & 45-50  & Male   & ChatGPT, Canva AI                      & Rarely \\
T9  & 35-40  & Male   & ChatGPT, Deepseek, Canva AI            & Always \\
T10 & 35-40  & Male   & ChatGPT, Canva AI                      & Always \\
\hline
\end{tabular}

\caption{Participant demographics with frequency of usage and locations. C refers to youth participants, P refers to parent participants, and T refers to teacher participants. All interviews were conducted individually, except for C1, C2, and P1, who were interviewed jointly at the parent's request.}
\label{tab:demo}
\end{table*}

\begin{table*}[ht]
\centering
\small
\setlength{\tabcolsep}{6pt}

\begin{tabular}{%
  >{\raggedright\arraybackslash\bfseries}p{0.16\linewidth}
  >{\raggedright\arraybackslash}p{0.38\linewidth}
  >{\raggedright\arraybackslash}p{0.38\linewidth}
}
\hline
\textbf{Topic} &
\textbf{Our Results} &
\textbf{Prior Work} \\
\hline

ChatGPT's role in religious interaction &
Saudi participants expressed serious concerns about GenAI shaping youth's religious knowledge, a role they believe belongs to parents. &
Not explored in Western studies of youth's GenAI use \cite{yu2025understanding,yu2024exploring}, nor in Middle Eastern social media contexts \cite{Abokhodair2016PrivacyS,AlSaggaf2016AnES}. \\

AI influence on youth's values \& perceptions &
Saudi participants worried about GenAI influencing youth's religious beliefs and cultural values. &
Western studies \cite{yu2024exploring} focused on employment concerns. Saudi parents previously expressed concerns about social media influencing youth's religion and thinking \cite{Alqahtani2017InternetRF,Moubarak2024RiskAO}. \\

Privacy vs.\ utility tension &
Saudi participants balanced traditional privacy expectations with GenAI utility (e.g., full name is private; age may be disclosed when benefits outweigh risks). &
Prior Saudi social media studies \cite{Abokhodair2016PrivacyS,AlSaggaf2016AnES} showed similar benefit-weighing attitudes toward public account disclosures. \\

Youth's trust in ChatGPT &
Saudi youth trusted GenAI highly, sometimes more than family, for current events, Islamic knowledge, and personal matters. &
Prior work \cite{yu2025understanding,yu2024exploring} found youth rely on GenAI for emotional support and develop parasocial relationships. \\

Shared/joint ChatGPT accounts &
Several Saudi users shared one subscription among 5+ people, raising concerns about inadvertent disclosure of private information, photos, and medical data. &
Novel finding in GenAI contexts. Related work on South Asian populations noted privacy concerns around shared devices \cite{10.1145/3544548.3581498}. \\

Sharing private information &
Parents worried youth would disclose full name, national ID, location, age, health data, phone number, photos, and others' private information. &
Both Western and non-Western parents share these concerns across digital platforms \cite{yu2024exploring,yu2025understanding,alashwali2022saudi}. \\

Phone content, microphone \& camera access &
Participants worried about GenAI accessing government ID apps (e.g., Absher) and ID photos; youth easily enabled camera/microphone in ChatGPT. &
Saudi parents refused or uninstalled apps requiring camera, microphone, or location access, citing risks to the entire family \cite{alashwali2022saudi}. \\

Boundaries for youth's GenAI use &
Safe-use boundaries included avoiding religion, morals, politics, and romantic/companion interactions with GenAI. &
Western parents similarly restricted topics such as history and news \cite{yu2025understanding,yu2024exploring}. \\

Expected controls \& safe GenAI responses &
Parents wanted control over youth's questions, culturally appropriate responses, and clear definitions of safe GenAI output. &
Western parents sought age-appropriate responses and topic-control features \cite{yu2025understanding,yu2024exploring}; both groups desired input moderation and topic oversight. \\

\hline
\end{tabular}

\caption{Summary of findings in comparison with prior work.}
\label{tab:findings}
\end{table*}

\appendix
\section{Interview Script}

Below is the full set of interview questions referenced in the Method section, organized by participant group.

\subsection{Introduction}
Hello, nice to meet you. My name is [Researcher Name], and I am a PhD student at Pennsylvania State University. Our group is conducting this interview to better understand [children's/parents'/teachers'] current use of generative AI and their views on privacy and safety when using Generative AI. Your honest responses are invaluable to our research, and all information provided will remain confidential.

\subsection{Interview with Youth}

The interview with youth was divided into three sections to answer our research questions.

\textbf{Section 1: Current State of Knowledge about GenAI}
\begin{itemize}
\item Do you know about Generative AI? 
\item What do you know about it? 
\item What GenAI tools have you used?
\item How often do you use these GenAI tools?
\end{itemize}

\textbf{Section 2: Experience and Use of GenAI Tools}
\begin{itemize}
\item Who do you learn about GenAI from?
\item Have your parents taught you about using GenAI tools?
\item What did they teach you? (follow-up)
\item Can you think of some recent uses of GenAI tools?
\item How did you use GenAI tools for that purpose? (follow-up, to elaborate on specific uses)
\end{itemize}

\textbf{Section 3: Privacy and Safety Practices in GenAI}
\begin{itemize}
\item What have your parents told you about privacy when using GenAI tools?
\item What have your teachers told you about privacy when using GenAI tools?
\item Could you tell me what kind of information you think you should keep private when using GenAI tools?
\begin{itemize}
\item Why do you think this information should be kept private? (follow-up)
\end{itemize}
\item Do you share GenAI accounts with your parents?
\begin{itemize}
\item If yes, could you tell me more about your experience sharing accounts with your parents?
\item Can you think of any benefits or risks of using a shared account with your parents?
\end{itemize}
\item Do you share your GenAI account with your siblings?
\begin{itemize}
\item If yes, could you tell me more about your experience sharing accounts with siblings?
\item Can you think of any benefits or risks of using a shared account with your siblings?
\end{itemize}
\item Do you share your GenAI account with others, such as friends or strangers?
\begin{itemize}
\item If yes, could you tell me more about your experience sharing accounts with others?
\item Can you think of any benefits or risks of using a shared account with others?
\end{itemize}
\item From your understanding, what are some ways you can safely use GenAI tools?
\begin{itemize}
\item Did you figure out safe use of GenAI tools by yourself, or did someone provide guidelines? If so, who?
\item Have your parents used any parental controls for your Generative AI tool account? Can you tell me more about it?
\item What would you like to learn about privacy and safety in GenAI?
\end{itemize}

\end{itemize}

\subsection{Interview with Parents and Teachers}

\textbf{Section 1: Experience and Use of GenAI Tools}
\begin{itemize}
\item Do you know about Generative AI? (warm-up)
\item What do you know about it? (follow-up)
\item What are some purposes you use a generative AI tool for?
\item How did you use GenAI tools for [tutoring] purposes? (based on response)
\item Does your [child/student] use GenAI tools? (warm-up)
\item What was the purpose? (follow-up)
\end{itemize}

\textbf{Section 2: Understanding and Addressing Privacy and Safety in GenAI}
\begin{itemize}
\item Do you know about privacy in AI?
\item Have you discussed the security and privacy risks of using GenAI tools with your [child/student]? (warm-up)
\item What did you discuss? (follow-up)
\item What safety and privacy concerns do you have regarding youth's use of GenAI tools?

\item Have you used any parental controls for your child Generative AI tool account? Can you tell me more about it?
\end{itemize}

\textbf{Section 3: Expectations for Improving Privacy and Safety}
\begin{itemize}
\item What would you like your [child/student] to learn about privacy and safety when using GenAI tools?
\item Who do you think is responsible for teaching these topics to youth?
\end{itemize}

\label{app:interview-script}

\label{app:interview-script}

\end{document}